\documentclass[aps,twocolumn,preprintnumbers,%
superscriptaddress,floatfix]{revtex4}

\usepackage[dvips]{graphics}
\usepackage{amsmath,amsfonts,bm}
\usepackage{psfig}

\begin{document}
\newcommand{\be}{\begin{eqnarray}}
\newcommand{\ee}{\end{eqnarray}}
\def\lsim{\mathrel{\rlap{\lower3pt\hbox{\hskip1pt$\sim$}}
     \raise1pt\hbox{$<$}}} 
\def\gsim{\mathrel{\rlap{\lower3pt\hbox{\hskip1pt$\sim$}}
     \raise1pt\hbox{$>$}}} 
\def\N{${\cal N}\,\,$}
\newcommand\<{\langle}
\renewcommand\>{\rangle}
\renewcommand\d{\partial}
\newcommand\LambdaQCD{\Lambda_{\textrm{QCD}}}
\newcommand\tr{\mathrm{Tr}\,}
\newcommand\+{\dagger}
\newcommand\g{g_5}
\def\la{\langle}\def\ra{\rangle}

\title{The Vector Manifestation of Hidden Local Symmetry, Hadronic Freedom\\ and
the STAR $\rho^0/\pi^-$ Ratio}
\author {G.E. Brown}
\affiliation { Department of Physics and Astronomy\\ State
University of New York, Stony Brook, NY 11794-3800}
\author{Chang-Hwan Lee}
\affiliation{Department of Physics, Pusan National University,
              Pusan 609-735 and
              \\
Asia Pacific Center for Theoretical Physics, POSTECH, Pohang
790-784, Korea}
\author{Mannque Rho}
\affiliation{ Service de Physique Th\'eorique,
 CEA Saclay, 91191 Gif-sur-Yvette c\'edex, France}
\begin{abstract}

The ``vector manifestation (VM)" fixed-point scenario found in
Harada-Yamawaki hidden local symmetry effective theory makes a
striking prediction that as the fire-ball expands and cools from
$T_c$ down toward the freeze-out temperature $T_{freezeout}$ in
heavy ion collisions, what we referred to as ``hadronic freedom"
sets in in which hadrons do not interact or barely interact. We
suggest that this scenario is testable at RHIC and as a specific
example, we provide an astonishingly simple prediction in terms of
the vector manifestation for the $\rho^0/\pi^-$ ratio measured in
peripheral collisions. We suggest that this hadronic freedom has
already been observed in the STAR peripheral collisions, in which
the $\rho^0$-mesons were reconstructed from following their pion
decay products back to the vertex where they are formed. In this
scenario, the measured $\rho^0/\pi^-$ ratio can be reproduced in
these special circumstances only because the ``flash temperature"
$T_{flash}$ at which the $\rho$-mesons go on-shell is equal to the
freeze-out temperature. In more central collisions with lower
freeze-out temperature, the rescattering of the pions is expected to
destroy the possibility of this reconstruction.
\end{abstract}


\newcommand\sect[1]{\emph{#1}---}

\maketitle

\sect{Introduction}%
The hidden local symmetry (HLS) effective theory, Wilsonian matched
to QCD at a matching scale $\Lambda_M$ near the chiral scale $\sim
4\pi f_\pi$, formulated by Harada and Yamawaki~\cite{HY,HY:PR},
predicts in the chiral limit that as the chiral restoration point
$P_c$ (where $P_c$ is the critical temperature $T_c$ or the critical
density $N_c$) is approached from below, the HLS coupling constant
$g$ as well as the vector-meson mass $m_V$ go to zero proportionally
to the quark condensate $\la\bar{q}q\ra$. Furthermore in heat bath
close to the critical temperature, the width $\Gamma$ goes to zero
even faster than the mass, going as $\sim \la\bar{q}q\ra^n$ with
$n\geq 2$. The original VM scenario of \cite{HY,HY:PR} was obtained
in a HLS theory that contained only the $\rho$ mesons and the pions
as relevant degrees of freedom but recently the argument has been
extended, independently by two groups, Harada and
Sasaki~\cite{HS-a1} and Hidaka et al~\cite{Hidakaetal}, to a
generalized HLS theory (GHLS) that incorporates the $a_1$ mesons
together with the $\rho$ and $\pi$ . The conclusion arrived at was
that although details differ, in all cases, the gauge coupling and
the mass for both vector and axial-vector mesons vanish as in the VM
scenario as the $P_c$ was approached. We refer to this as
``generalized vector manifestation (GVM)" scenario.

In \cite{BGR}, the notion of ``hadronic freedom" in the vicinity of
$P_c$ was formulated in terms of the VM (or GVM). Near the VM, due
to the vanishing gauge coupling, hadronic interactions become weak,
vanishing at the critical point. This suggests that processes that
take place near $P_c$ can be described more efficiently by
fluctuating around the VM fixed point than around the matter-free
vacuum as is customarily done by the workers in the field. This
notion has been recently applied to kaon condensation in compact
stars.  See \cite{BLR-kaon}. In this paper, we will exploit this
idea and make a remarkably simple prediction for the ratio
$\frac{\rho^0}{\pi^-}$ in peripheral heavy ion collisions and
compare with what had been obtained by the STAR
collaboration~\cite{star}.

What makes our calculation particularly simple, allowing us to
bypass a plethora of possible complications, is that the ``flash"
temperature (defined below) and the ``freeze-out" temperature
coincide in the peripheral collision kinematics of the STAR
experiment. The two temperatures are different in more central
collisions in which case the problem becomes considerably more
involved. We will further elaborate on this point in what follows.

\sect{The STAR experiment} In a recent experiment, STAR~\cite{star}
has reconstructed the $\rho$-mesons from the two-pion decay products
in the Au + Au peripheral collisions at $\sqrt{s_{NN}}= 200$ GeV.
They find the ratio of
 \be
\frac{\rho^0}{\pi^-}|_{STAR}=0.169\pm 0.003 ({\rm stat})\pm 0.037
({\rm syst}),\label{stardata}
 \ee
almost as large as the $\rho^0/\pi^-=0.183\pm 0.001 ({\rm stat})\pm
0.027 ({\rm syst})$ in proton-proton scattering. The near equality
of these ratios was not expected, in that the $\rho$ meson width
$\Gamma\sim 150$ MeV in free space is the strongest meson
re-scattering that one has. For instance, at SPS one needs about 10
generations in order to get the dileptons~\cite{rapp}. Furthermore,
if one assumes equilibrium at the freezeout, then the ratio is
expected to come to several orders of magnitude smaller, say,
$\frac{\rho^0}{\pi^-}\sim 4\times 10^{-4}$~\cite{peteretal}.

We will perform our calculation in kinematic conditions that we take
to be those met in the STAR experiment. Assuming that the $\rho^0$
reconstruction is made at the ``flash point" $T_{flash}$ (defined
precisely below) after only one generation from $T_c$ and that the
hadronic freedom, together with Brown-Rho scaling~\cite{BR}, is
operative to that point, we find
 \be
0.16\lsim \frac{\rho^0}{\pi^-}\lsim 0.21.\label{range}
 \ee
In what follows, we describe how we arrive at this result.

\sect{The vector manifestation and ``hadronic freedom"}%
We start by briefly describing the theoretical framework employed in
our work, namely, the hidden local symmetry approach. In the context
of strong interactions, the structure of hidden local symmetry
arises naturally when one attempts to go up systematically in scale
from low-energy chiral Lagrangians of Goldstone bosons to the regime
where the (Goldstone-only) effective theory breaks down due to the
onset of new (heavier) degrees of freedom~\cite{georgi}. One
approach where local gauge invariance emerges in this way is the
``moose" construction~\cite{georgi} to model QCD that leads to a
dimensional deconstruction of a 5-D Yang-Mills Lagrangian which
describes low-energy hadronic dynamics in terms of an infinite tower
of 4-D hidden gauge bosons gauge coupled to Goldstone bosons that
appear as the Wilson line over the fifth component of the 5-D gauge
field~\cite{son-stephanov}. The infinite tower of hidden gauge
bosons also arise from holographic dual AdS/QCD approach to strong
interactions in which 5-D gravity theory which compactified, gives
rise to 4-D hidden local symmetry theory~\cite{sakai-sugimoto}. The
local gauge invariance in the 5-D bulk theory reflects the global
chiral invariance in the 4-D gauge theory QCD which is made
gauge-equivalent to hidden local invariance in 4-D. These bulk
Lagrangians (in the gravity sector) are in a manageable form only
for large $N_c$ limit (more precisely 't Hooft limit) and hence
physical quantities are evaluated in terms of tree graphs, i.e.,
mean field. Unfortunately at present there is no fully systematic
way to compute the corrections to the large $N_c$ which are expected
to be important for physics near the critical point. Indeed, a
recent work by Harada et al~\cite{HMY} who developed a technique to
calculate the leading $1/N_c$ (one-loop) corrections to the
holographic dual AdS/QCD model of \cite{sakai-sugimoto} in doing
chiral perturbation theory shows that leading-order corrections
could be important for certain $\rho$-meson properties.

The HLS effective field theory approach of Harada and
Yamawaki~\cite{HY} instead focuses on the lowest member of the tower
and incorporates leading quantum corrections at one-loop order which
play a crucial role in the context we are interested in. To do this,
Harada and Yamawaki cut off the theory at a scale $\Lambda_M$ (near
the chiral scale $\Lambda_\chi\sim 4\pi f_\pi$) that lies above the
lowest member of the tower, $\rho$, but below the next member,
$a_1$,  make the Wilsonian matching to QCD at $\Lambda_M$ and
compute quantum corrections with the bare Lagrangian so defined with
the {\it intrinsic background dependence} duly incorporated at the
scale $\Lambda_M$. The Wilsonian matching effectively takes into
account the physics associated with the degrees of freedom lying
above the matching scale, including $a_1$, glueballs etc. They find
that the theory has a fixed point {\it consistent with chiral
symmetry of QCD} at which the $\rho$ mass and the vector coupling
constant $g$ go to zero. This is the vector manifestation (VM) fixed
point. It has been further shown that the VM fixed point is reached
when matter is heated to some critical temperature $T_c$~\cite{HS-T}
or compressed to some critical density $n_c$~\cite{HKR} or if the
number of flavors reaches a critical value $N_f^c \gsim
5$~\cite{HY}.

The most important element in the HLS approach for our discussion to
be developed below is that it predicts that the vector-meson mass
goes to zero near the VM fixed point $because$ the gauge coupling
$g$ goes to zero. That $g$ goes to zero follows from the
renormalization group flow for the gauge coupling $g$. At one loop
-- higher loops are unaccessible at present~\cite{footnotloops}, the
gauge coupling runs with the beta function that is of the form
 \be
 \beta(g)=- \frac{N_f}{2(4\pi)^2} \frac{87 - a^2}{6} g^4
 \ee
where $N_f$ is the number of flavors and $a=F_\sigma^2/F_\pi^2$ with
$F_\sigma$ the decay constant of the scalar that gets higgsed to
make $\rho$ massive and $F_\pi$ the pion decay constant. The
parameter $a$ also runs, so the flow structure of the coupled RGEs
is more complicated than in QCD but $a$ is near 1, so the beta
function is always negative. This means that $g$ has an
``ultraviolet fixed point" $g=0$, that is, the theory is in a sense
asymptotically free~\cite{footnoteAF}. Thus as one approaches the
critical point, $g$ approaches its asymptotically free value. Since
the interactions between hadrons are largely governed by this
coupling constant, we refer to this phenomenon as ``hadronic
freedom." What this implies is that right below the critical point,
there is a region where hadronic interactions are absent or
negligible. Clearly this would make the scenario for heavy-ion
dynamics drastically different from the standard (linear sigma
model) scenario adopted by the heavy-ion community~\cite{BGR,BLR05}.

In order to implement VM/HLS in heavy-ion dynamics relevant to the
STAR data, we need to introduce additional degrees of freedom to the
$\pi$ and $\rho$ figuring in the original Harada-Yamawaki theory
below the matching scale $\Lambda_M$. It turns out that this can be
done for the $a_1$. In a recent important work, two independent
groups, Harada and Sasaki~\cite{HS-a1} and Hidaka et
al~\cite{Hidakaetal}, have incorporated the $a_1$ meson into hidden
local symmetry framework and shown that the VM-like fixed point
(say, a generalized VM fixed point) exists at which the gauge
coupling $g$ {\it as well as} the $\rho$ and $a_1$ masses go to
zero. The fixed point structure is a bit more complicated than
Harada-Yamawaki's VM. However for our purposes, this complication is
of no consequence. We shall therefore continue our discussion using
the VM/HLS without losing validity for the extended fixed point
structure. Thus far, scalar mesons  -- which also figure in the
counting of the relevant degrees of freedom -- have not been
incorporated. However there are two reasons to believe that the
scalar mesons also join the vectors at the fixed point, going
massless at that point with vanishing interactions. One is that the
multiplet structure implicit in the calculations of
\cite{HS-a1,Hidakaetal} is such that the scalar (``$\sigma$") meson
joins either the $a_1$ or $\pi$, massless in the chiral limit.
Another is Weinberg's mended symmetry for certain helicity-zero
states~\cite{weinberg} which suggests that at $T_c$, $\pi$,
$\epsilon$, $\rho$ and $a_1$ become massless with the vector and
axial vector mesons becoming local gauge bosons.

We must admit that while the VM scenario (in the form of Brown-Rho
scaling) has been found to be generally consistent with a variety of
observables with no evidences against it~\cite{review-BR} and
furthermore somewhat surprisingly the VM/HLS fixed point is proven
to be a good starting point even for certain processes taking place
far away from the critical point~\cite{HRS} (see \cite{rho-nagoya}
for review), there has been no rigorous theoretical proof for or
against it at the critical point from QCD $proper$. Lattice
measurement of the vector meson pole mass at $T_c$ would be a
crucial test but this is not yet available. On the experimental
side, there are now unambiguous evidences for Brown-Rho scaling at
near nuclear matter density, e.g., ~\cite{omega}. However there has
been up to date no direct experimental indication for VM near the
critical point, temperature or density. We believe that the STAR
$\rho^0/\pi^-$ ratio could offer an experimental test for the VM (or
GVM) scenario.

\sect{Flash temperature $T_{flash}$} A preliminary model to
understand the STAR ratio (\ref{stardata}) in terms of
Harada-Yamawaki theory was given in Appendix C of Ref. \cite{BLR}
where it was suggested that due to the VM, the $\rho$ meson was
prevented from decaying into two pions until thermal freezeout. Our
conclusion there was that essentially no $\rho$-mesons decay for
about 5 fm/c as the temperature drops below $T_c$ into what is
usually called ``mixed phase," and then they decay at a temperature
that we shall refer to as the ``flash temperature" $T_{flash}$ since
nearly all the $\rho$ and $a_1$ mesons decay at this temperature --
the latter as $a_1\rightarrow \pi + \rho \rightarrow 3\pi$. We will
note later that $T_{flash}\simeq T_{freezeout}$ (where
$T_{freezeout}$ is the freezeout temperature) for peripheral
collisions, so that the pions coming from the vector-meson decay are
able to leave the fireball without re-scattering.

We will now simplify and sharpen the model introduced in \cite{BLR}
and make a quantitative prediction for the ratio.

Let us first define $T_{flash}$ of the STAR data following the work
of Shuryak and Brown~\cite{SB}. Shuryak and Brown determined the
flash temperature to be the temperature at which the $\rho$ was 90\%
``on-shell" (in free space), namely, when the $\rho$ mass is 700
MeV. Some of the 10\% decrease came from the Boltzmann factors on
the colliding particle producing the $\rho$ but the shift of - 28
MeV came from Brown-Rho scaling for a scalar density of $\sim$ 0.15
$n_0$ ($n_0$ is nuclear matter density) and -10 MeV came from the
scalar density of other vector mesons. Since the $\rho\rightarrow
2\pi$ is p-wave, there should be a ``kinematical" effect in which
the negative mass shift substantially reduces the width, both
because of the reduced phase space and also due to the presence of
the momentum in the p-wave matrix element. The magnitude of the
effect for the 10\% mass shift is $ \delta\Gamma_\rho\approx
3\frac{\delta m_\rho}{m_\rho}\approx -50\ {\rm MeV}$. This leads us
to believe that the STAR detector measured a $\rho$-meson 90 \%
on-shell with a mass of 700 MeV and a width 2/3 of the free space
value.~\cite{footnote1}

Whereas the flash temperature is independent of the centrality,
this is not the case for the freezeout temperature
$T_{freezeout}$. In fact $T_{freezeout}$ decreases as the
centrality increases.  Thus, if, as we find, $T_{freezeout} \simeq
T_{flash}$ for peripheral collisions, $T_{freezeout} < T_{flash}$
for central collisions and in the time the system is between these
two temperatures the pions from $\rho$ and $a_1$ decay will be
rescattered as it is no longer possible from their detection to
work backward to the parent vector mesons.

It is possible to pin down the numerical value of the flash
temperature from lattice calculations. We shall extract this
quantity from Miller's lattice calculation~\cite{miller} which we
analyzed in \cite{BLR05}. The result is
 \be
T_{flash}\approx 120\ {\rm MeV}.
 \ee
How this result comes about is as follows.   As pointed out in
\cite{BLR05}, in Miller's lattice calculation of the gluon
condensate, the soft glue starts to melt at $T\approx 120$ MeV. The
melting of the soft glue, which breaks scale invariance as well as
chiral invariance $dynamically$ -- and is responsible for Brown-Rho
scaling, is completed by $T_c$ at which the particles have gone
massless according to the VM. One can see that the gluon condensate
at $T\sim 1.4\ T_c$ is as high as that at $T\sim T_c$. This
represents the hard glue, or what is called ``epoxy," which breaks
scale invariance $explicitly$ but has no effect on the hadron mass.
We see that the melting of the soft glue is roughly linear, implying
that the meson masses drop linearly with temperature. This
connection between the melting of the soft glue and the dropping of
the meson mass has recently been confirmed in QCD sum-rule
formalism~\cite{BHR}.

\sect{Calculation of the STAR ratio} In describing what happens as
the system expands and the temperature cools from $T_c$ down to
$T_{flash}$, we choose the 32 degrees of freedom for the $\rho$,
$\pi$, $\sigma$ and $a_1$ in the $SU(4)$ multiplet (for up and down
flavors) that come down from above $T_c$ as described in
\cite{BLR05,BGR}. These are the light degrees of freedom found in
the quenched lattice calculation of Asakawa et
al.~\cite{asakawaetal} and used in \cite{BLRS}. We suggest that the
entropy at $T_c$ in lattice calculations correspond to that of
massless bosons. Below $T_c$, the hadronic freedom is operative with
the mesons nearly massless (not strictly massless due both to the
non-zero quark mass and to $T\neq T_c$). We assume that the pion
which as pseudo-Goldstone has a small non-zero mass, stays more or
less constant as temperature drops from $T_c$ to $T_{flash}$. Now in
Harada-Yamawaki HLS/VM, near $T_c\approx 175$ MeV, the width drops
rapidly as the mass drops. Including the p-wave penetration factor
as in \cite{SB} would make it drop as fast as
 \be
\frac{\Gamma^\star_\rho}{\Gamma_\rho} \sim
\left(\frac{m_\rho^\star}{m_\rho}\right)^3
\left(\frac{g^\star}{g}\right)^2 \Rightarrow
\left(\frac{\la\bar{q}q\ra^\star}{\la\bar{q}q\ra}\right)^5.
 \ee
We assume that the effective gauge coupling in medium denoted
$g^*$ begins to scale only above $T_{flash}\approx
T_{freezeout}\approx 120$ MeV in peripheral collisions when the
soft glue has begun to melt. This is analogous to the behavior in
density where it is empirically established that the scaling of
$g^*/g$ sets in only at $n\sim n_0$~\cite{BR2004}.

\begin{table}
\caption{$\Gamma^\star$ as function of temperature. For the point
at 120 MeV we have switched over to the Shuryak and Brown
\cite{SB} value for $\Gamma^\star/\Gamma$. } \label{tab1}
\begin{center}
\begin{tabular}{ccc}
\hline
T  & $m_\rho^\star/m_\rho$ & $\Gamma^\star/\Gamma$ \\
\hline
175 MeV &  0 & 0 \\
164 MeV & 0.18 & 0 \\
153 MeV & 0.36 & 0.01 \\
142 MeV & 0.54 & 0.05 \\
131 MeV & 0.72 & 0.22 \\
120 MeV & 0.90 & 0.67 \\
\hline
\end{tabular}
\end{center}
\end{table}

Our next task is, then, to reverse the process by which the mesons
lost their masses by the melting of the soft glue as $T$ increased
from 120 MeV to the chiral restoration temperature $T_c=175$ MeV,
in order to show how the mesons get their masses back; i.e., go
on-mass-shell in free space.

In Table \ref{tab1} is shown the calculated evolution of the mass
and width of the $\rho$ as the temperature drops from $T_c$. At
$T\sim 120$ MeV which is the temperature at which the soft glue
begins to melt and the $\rho$ is 90\% on-shell, the $\rho$ meson
decays into two pions. Down to that temperature, the width $\Gamma
(\rho\rightarrow 2\pi)$ is too small to do much before then. The
meson has to be nearly on-shell before it can decay. Once
on-shell, it decays very rapidly. The same argument holds for
$a_1$ and other mesons in the $SU(4)$ multiplet.

We find that in total 66 pions result at the end of the first
generation from the 32 $SU(4)$ multiplet, i.e., $\rho$ (18), $a_1$
(27), $a_0$ (4), $\pi$ (3), $\sigma$ (2) and $\epsilon\equiv f
(1285)$ (12) where the number in the parenthesis is the number of
pions emitted. Excluded from the counting are the $\omega$ and
$\eta$ since they leave the system before decaying. Leaving out the
three $\pi^-$'s coming from the $\rho^0$ decays which are
reconstructed in the measurement, we obtain
  \be
\frac{\rho^0}{\pi^-}\approx 3/(22-3)\approx
0.16.\label{prediction}
 \ee
Of course not all of the vector and axial vector mesons will have
decayed. There could be a few left over. On the other hand some
pions will come from the $\pi$, $\sigma$ group of initial degrees
freedom.  Now the above estimate of 0.16 for the
$\frac{\rho^0}{\pi^-}$ ratio assumed all vector and axial-vector
mesons decayed into pions at the flash temperature. This means that
(\ref{prediction}) should be taken as a lower limit. If we take the
numbers in Table \ref{tab1} literally and assume a dynamical time of
1 fm/c, we find that only 42 of the 66 total number of pions have
been emitted at $T\approx 120$ MeV. If we take 42/3 $\pi^-$-mesons,
we have
 \be
\frac{\rho^0}{\pi^-}\approx 0.21 \label{prediction2}
 \ee
which we can take as an upper limit. Thus we arrive at the range of
our prediction announced in (\ref{range}). This is in a surprising
accord with the observed value (\ref{stardata}) considering that the
standard scenario would be off by several orders of magnitude.

The factor of $\gsim 400$ enhancement with respect to the
equilibrium value can be understood as follows. Because of the
decreased width due to the hadronic freedom, the $\rho$ meson goes
through {\it only one generation} before it freezes out in the
peripheral collisions in STAR. But there is no equilibrium at the
end of the first generation. Had the $\rho$ possessed its on-shell
mass and width with $\sim$ five generations as in the standard
scenario, it is clear that the $\frac{\rho^0}{\pi^-}$ ratio would be
much closer to the equilibrium value.

\sect{HBT} The scenario proposed in this paper has a simple
implication on HBT. The roughly 5 fm/c expansion without
hadronization sharply distinguishes the Harada-Yamawaki theory
from those used to date. In particular such a large expansion will
give a balloon-shaped fireball, which serves to make the outwards
bound, sideways flow and longitudinal HBT look quite similar, as
observed.

\sect{Conclusion} Adopting the extended vector manifestation for the
$\rho$ and $a_1$ mesons~\cite{HS-a1,Hidakaetal} and assuming that
the scalar mesons join the vectors at the fixed point and that the
resultant hadronic freedom is operative from $T_c$ down to $\sim
T_{flash}$ at which both the $\rho$ and $a_1$ decay to pions, we
predict a surprisingly large value for the STAR ratio in contrast to
the standard scenario where the ratio is expected to be much
smaller. With nearly zero couplings and masses, the particles stream
freely without interaction until the vector mesons go about 90\%
on-shell at the flash temperature $T_{flash}\approx 120$ MeV at
which the soft glues condense and induce strong interactions
triggering the decay into pions.

At present no calculations in Harada-Yamawaki VM/HLS theory are
available to indicate how the coupling constant (and consequently
the mass) of the $\rho$-meson  evolves as the temperature goes
down from $T_c$. On the other hand, if the argument we developed
in this paper is correct, then the STAR ratio is telling us that
nothing much happens to the coupling constant until the
temperature reaches $T_{flash}$.

We note that how the hadronic system evolves in the vicinity of
chiral restoration (or deconfinement) is an important theoretical
issue in QCD as it concerns the basic structure of the QCD vacuum.

\sect{Acknowledgments} GEB is grateful for enlightening
discussions on the STAR data with Edward Shuryak and MR would like
to thank Masa Harada and Chihiro Sasaki for tuition on VM. The
work of GEB was supported in part by the US Department of Energy
under Grant No. DE-FG02-88ER40388 and that of CHL by the grant No.
R01-2005-000-10334-0(2005) from the Basic Research Program of the
Korea Science \& Engineering Foundation.


\begin{thebibliography}{99}
\bibitem{footnoteAF} The term ``asymptotic freedom" as used here
could be a misnomer in that the theory we are using is an effective
theory defined up to the cutoff at the matching scale whereas QCD
which is asymptotically free is a renormalizable theory where the
cutoff can be sent off to infinity. What it means is that within the
scale applicable to the effective theory, there is an ultraviolet
fixed point at which the coupling constant goes to zero.
\bibitem{footnote1} Empirically
the total width of the $\rho$ measured turned out to be $\sim$ 150
MeV, not 100 MeV, because of the collisional broadening of $\sim$
50 MeV.
\end{thebibliography}
\end{document}